\documentclass{article}
\usepackage[utf8]{inputenc}

\title{Analytical solution of a plane strain pure bending problem in second gradient electroelasticity}
\author{Yury Solyaev, \quad Sergey Lurie}
\date{\textit{Institute of Applied Mechanics of the Russian Academy of Sciences, Moscow, Russia} \\[5pt] 
email: yos@iam.ras.ru \\[10pt] 
\today}

\usepackage{graphicx}
\usepackage[numbers]{natbib}
\usepackage[misc,geometry]{ifsym}
\usepackage{amsmath}
\usepackage{amsfonts}
\usepackage{bm,nicefrac}
\usepackage{amsfonts}
\usepackage{enumitem}
\usepackage{siunitx}
\usepackage[caption=false]{subfig}
\usepackage{hyperref}

\begin{document}

\maketitle

\begin{abstract}
Semi-inverse analytical solution of a pure bending problem for piezoelectric layer is developed in the framework of linear electroelasticity theory with strain gradient and electric field gradient effects. Two-dimensional solution is derived assuming plane strain state of a layer. It is shown that obtained solution can be used for the validation of size-dependent beam and plate models in second gradient electroelasticity theory.\\ 
\end{abstract}

\section{Introduction}
\label{sec:1}
Size-dependent behavior of piezoelectric structures attracts high attention in last years due to its possible importance for novel micro- and nanoelectromechanical systems such that sensors, energy harvesters, nanopositioners etc. \cite{Yang2006, majdoub2008, Yan2017, Oates2017}. Different non-classical theories of piezoelectric materials have been proposed to explain and to predict new experimental results taking into account flexoelectricity \cite{Sharma2007, Tagantsev2013, Wang2018}, polarization field and electric field gradients \cite{mindlin1968, sahin1988, kafadar1971, Arvanitakis2018}, stress and strain gradients \cite{Liang2013x, Enakoutsa2015, Iesan2018, Liu2016}, couple stresses \cite{Hadjesfandiari2013, Malikan2017}, surface effects \cite{liang2014, shen2010}. Corresponding beam and plate theories were developed to explain non-classical dependence of apparent electromechanical properties of thin piezoelectric cantilevers and membranes on its size (see e.g. \cite{majdoub2008, Wang2012, liang2013, wang2018b, yue2015b, yang2015b, Baroudi2018}). However, for the best of the authors knowledge, despite some experimental tests and molecular dynamics simulations \cite{majdoub2008}, theoretical validation of these simplified models has not been presented yet. Nevertheless, it is well known that comparison with exact three-dimensional or plane strain/plain stress solutions should be provided to demonstrate a correctness and strong theoretical background of established beam and plate models. 

For example, in classical elasticity validation of an Euler-Bernoulli beam model can be obtained by using Saint-Venant's problems solutions assuming high length-to-thickness ratio of a beam and neglecting of Poisson's effect \cite{lurie2005}. In classical piezoelectricity, Saint-Venant's solutions for piezoelectric cylinders have been also obtained \cite{Iesan1989, Davi1996, DellIsola1996, Bisegna1998, Rovenski2007}. These solutions together with finite-element simulations can be used to verify a classical theory of piezoelectric beams with different order of through-thickness approximation of the displacement and electric fields (see, \cite{Krommer2001} and references therein).  Plane strain and three-dimensional solutions for verification of classical piezoelectric plate models have been given in \cite{Heyliger1996, Heyliger1997}. 

Verification of non-classical beam and plate models in generalized continuum theories is of highly importance due to its complex nature and a lot of additional constitutive relations and boundary conditions. Such validation of an Euler-Bernoulli beam model in microdilatation elasticity \cite{Cowin1983} have been given in \cite{Birsan2011} by using semi-inverse solution of a beam pure bending problem. Validation of beam theories in micropolar and microstretch elasticity can be obtained by using known solutions of corresponding generalized Saint-Venant's problems \cite{iesan1971, reddy1978, iesan1994}. In the authors recent work \cite{Lurie2018, lurie2016nanosized}, validation of gradient beam bending theories in Mindlin Form I and Form II  \cite{Mindlin1964, mindlin1968b} have been presented based on three-dimensional semi-inverse solution of a beam pure bending problem. It was shown a significant importance of correct formulation of gradient beam models taking into account additional boundary conditions on the top and bottom surfaces of the beam that were ignored in a number of works. 

Subject of the present paper is a second gradient electroelasticity theory that takes into account an influence of strain and electric field gradients. This theory has been developed in \cite{Kalpakides2002, hu2009, Liang2013x}. Recently it was applied to the problems of fracture mechanics \cite{Sladek2018} and micromechanics \cite{Yue2014, Solyaev2018}. Beam-type models in this theory were developed in \cite{yue2015b} and with strain gradient effects only in \cite{liang2013, Liang2013x}. Similar theory with distortion gradient and thermal effects were presented recently in \cite{Iesan2018}. In the present paper we derive a semi-inverse analytical solution of a pure bending problem for piezoelectric layer that can be used for verification of corresponding beam and plate models in considered theory. We obtained analytical solution following an approach that were used in \cite{Lurie2018} generalizing it for piezoelectric effects. Due to complex form of second gradient electroelastictity theory, we consider static plane strain statement of a problem, i.e. cylindrical bending of a plate. For the same reason, simplified variant of constitutive equations with single additional length scale parameter is used following \cite{Yue2014, yue2015b}. Thus, in the absence of electromechanical coupling, considered theory can be reduced to the widely known simplified strain gradient elasticity theory \cite{askes2011, park2006}.

\section{Second gradient electroelasticity theory}
\label{sec:2}

Let us consider an electro-elastic body occupying the region $\Omega$ with boundary $\partial \Omega$. The electric Gibbs energy density in considered theory can be expressed in the following form \cite{hu2009, yue2015b, Yue2014}:
\begin{equation}
\label{g}
\begin{aligned}
	g(\varepsilon_{ij}, \varepsilon_{ij,k}, E_i, E_{i,j}) =
	 &\, \tfrac{1}{2} C_{ijkl} \varepsilon_{ij} \varepsilon_{kl} -
	 e_{kij} \varepsilon_{ij} E_k \, - \tfrac{1}{2} \kappa_{ij} E_i E_j +\\[2mm]
	 &\, \tfrac{1}{2} A_{ijklmn}\varepsilon_{ij,k} \varepsilon_{lm,n} - 
	\tfrac{1}{2}\alpha_{ijkl}E_{i,j}E_{k,l}
\end{aligned}
\end{equation}
where $C_{ijkl}$ and $A_{ijklmn}$ are the fourth- and sixth-order tensors of the elastic moduli; $e_{kij}$ is the third-order tensor of piezoelectric moduli; $\kappa_{ij}$ and $\alpha_{ijkl}$ are the second- and fourth-order tensors of dielectric permittivity constants; $\varepsilon_{ij}$ is an infinitesimal strain tensor and $E_i$ is electric field vector that are defined by the relations:
\begin{equation}
\label{u}
	\varepsilon_{ij} = \tfrac{1}{2}(u_{i,j}+u_{j,i})
\end{equation}
\begin{equation}
\label{E}
	\quad E_i = - \phi_{,i}
\end{equation}
where $\textbf{u}(\textbf{x})$ and $\phi(\textbf{x})$ are the vector of mechanical displacements and electric potential function, respectively, at a point $\textbf{x} = \{x_1, x_2, x_3\}$; the comma denotes the differentiation with respect to spatial variables and repeated indices imply summation.

Note, that in (\ref{g}) we neglect the flexoelectric effects and high-order coupling effects, including coupling between strain and strain gradient tensors that can persist in the general case of non-centrosymmetric piezoelectric materials. From one side we make these assumptions to consider a most simple gradient electro-elasticity theory because our goal is to derive a closed form analytical solution that could be used as the test solution for validation of beam and plate theories. From the other side, experience has shown that even simplified and reduced gradient theories can be useful in practical application \cite{eremeyev2017}. 

Following \cite{Yue2014, yue2015b}, we assume the relations between high-order and classical moduli: $A_{ijklmn} = C_{ijlm} \ell^2_{kn}$ and $\alpha_{ijkl} = \kappa_{ik} \rho^2_{jl}$, where $\ell_{kn}$ and $\rho_{jl}$ are the tensors of the length scale parameters that define the gradient effects in elastic and electrical fields, respectively. For simplification, we additionally assume that the length scale parameters are the same in elasticity and electrostatics problems and do not depend on the direction, i.e. $\ell_{ij} = \rho_{ij}= \ell^2 \delta_{ij}$. Thus, the model consists single additional length scale parameter $\ell$. In the case of $\ell = 0$, electric Gibbs energy density (\ref{g}) reduces to the classical form of electroelasticity theory \cite{parton1988}.

Constitutive equations of linear theory follow from (\ref{g}):
\begin{equation}
\label{s}
	\quad \sigma_{ij} = C_{ijkl} \varepsilon_{kl} - e_{kij} E_k
\end{equation} 
\begin{equation}
\label{t}
	\tau_{ijk} = A_{ijklmn} \,\varepsilon_{lm,n}
\end{equation} 
\begin{equation}
\label{D}
	D_i = e_{ijk} \varepsilon_{jk} + \kappa_{ij} E_j
\end{equation} 
\begin{equation}
\label{Q}
	Q_{ij} = \alpha_{ijkl} E_{k,l}
\end{equation} 
where $\sigma_{ij}$ is Cauchy stress tensor, $\tau_{ijk}$ is the third-order double stress tensor \cite{mindlin1968b}, $D_i$ is the electric displacement vector, $Q_{ij}$ is the electric quadrupole tensor \cite{kafadar1971, Yang2004}.

Field equations and boundary conditions of the model can be obtained based on variational approach (see \cite{hu2009, Yue2014, Iesan2018}). Assuming the absence of body forces and free charges one can obtain the equations of elastic equilibrium and generalized Gauss' law of electrostatics:
\begin{equation}
\label{Equil}
	\sigma_{ij,j} - \tau_{ijk,jk} = 0 \qquad \textbf{x}\in\Omega
\end{equation} 
\begin{equation}
\label{Gauss}
	D_{i,i} - Q_{ij,ij} = 0 \qquad \textbf{x}\in\Omega
\end{equation} 

Boundary conditions are prescribed on the body surface $\partial \Omega$ assuming absence of high-order loading (zero right parts in the equations (\ref{BCt})-(\ref{BCQ})) and have the following form:
\begin{equation}
\label{BCs}
\begin{aligned}
	\quad (\sigma_{ij} - \tau_{ijk,k}) n_j - (\tau_{ijk} n_k)_{,j} + 
	(\tau_{ijk} n_k n_l)_{,l} \,n_j = \overline{t}_i
	\quad or \quad
	u_i = \overline{u_i} 
\end{aligned}
\end{equation} 
\begin{equation}
\label{BCD}
\begin{aligned}
	(D_{j} - Q_{jk,k}) n_j - (Q_{jk} n_k)_{,j} + (Q_{jk} n_k n_l)_{,l} \,n_j = \overline{q}
	\quad or \quad
	\phi = \overline{\phi}
\end{aligned}
\end{equation} 
\begin{equation}
\label{BCt}
\begin{aligned}
	\tau_{ijk} n_j n_k = 0 \quad or \quad u_{i,j}n_j = 0
\end{aligned}
\end{equation} 
\begin{equation}
\label{BCQ}
\begin{aligned}
	\quad Q_{ij} n_i n_j = 0  \quad or \quad E_i n_i = 0
\end{aligned}
\end{equation} 
where $n_i$ is the vector of the external unit normal to $\partial\Omega$; $\overline{t_i}$, $\overline{u_i}$, $q$, $\overline{\phi}$ are the prescribed mechanical traction vector, displacement vector, surface electric charge density and electric potential, respectively.

Additional boundary conditions of a gradient theory should be prescribed at the sharp edges of the body $\Gamma(\partial\Omega)$ as follows:
\begin{equation}
\label{Edge}
	[\tau_{ijk} m_j n_k] = 0, \qquad [Q_{jk} m_j n_k] = 0
\end{equation} 
where $m_j= \epsilon_{jml}s_m n_l$ is the outer co-normal vector; $s_m$ is the unit vector tangent to the given edge; the brackets [...] indicate that the enclosed quantity is the difference between the values on the body faces to which the given edge belongs; $\epsilon_{jml}$ is the permutation symbol.

\section{Pure bending of piezoelectric layer}
\label{sec:3}

\begin{figure}[h]
\centering
\includegraphics[width=0.8\textwidth]{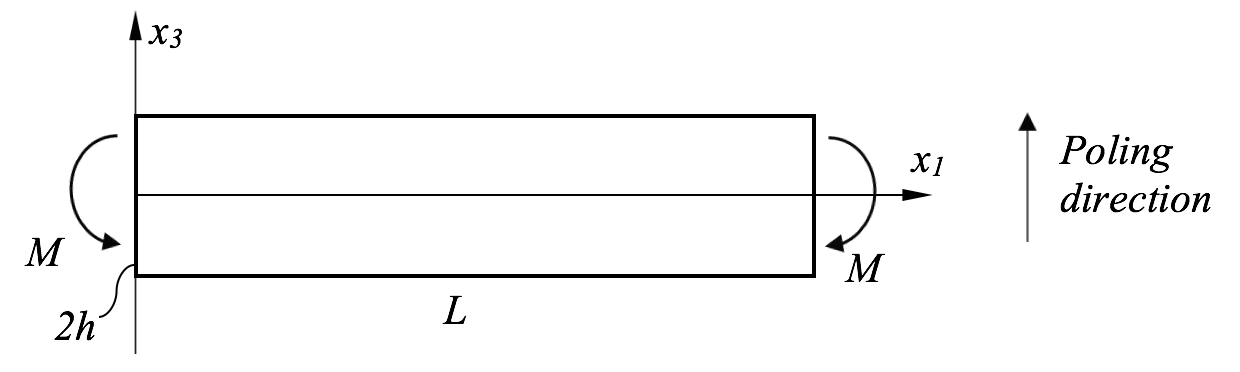}
\caption{Pure bending of piezoelectric layer}
\label{fig1}       
\end{figure}

Consider a piezoelectric layer of length $L$ and thickness $2h$, i.e. $\Omega = \{\textbf{x}: x_1\in[0,L], x_3\in[-h, h]\}$ (Fig. \ref{fig1}). Layer has infinite dimension in out-of-plane direction $x_2$, thus, a plane strain state hypothesis is valid: we assume that there are no displacements in transversal direction ($u_2=0$), all fields quantities do not depend on $x_2$ coordinate and boundary conditions on the lateral sides ($n_i=\{0,1,0\}$) can be neglected. The layer is poled along the thickness direction, so that it exhibits the material symmetry of a hexagonal crystal in class 6mm -- transversely isotropic about $x_3$ axis. 

Boundary conditions for the considered problem are the following. At the end faces of the layer $x_1 = 0$ and $x_1 = L$ the distributed loading $\overline{t_i} = \{\overline{t}_1(x_3),0,0\}$ is acted with resultant bending moments $M$ (distributed moments per unit width) and with zero force resultants, such that pure bending is realized in the $x_1x_3$ plane about the $x_2$ axis. There are no stresses acting on the top and on the bottom faces of the layer. The surface charge density $\overline{q}$ is zero over the whole boundary, thus, the electric permittivity of the surrounding medium (for example air) assumed to be much less than that of the piezoelectric layer. The layer is fixed at the origin of coordinate system, where the zero value of electric potential is also prescribed. 

For a given statement of the problem, the component representations of the boundary conditions (\ref{BCs})-(\ref{BCQ}) become:

\begin{equation*}
	\hspace{-3.5cm} \underline{x_1 = 0, L}:
\end{equation*}
\begin{equation}
\label{1}
	\sigma_{11} - \tau_{111,1} - \tau_{113,3} - \tau_{131,3} = 
	\overline{t}_1(x_3)
\end{equation}
\begin{equation}
\label{2}
	\sigma_{21} - \tau_{211,1} - \tau_{213,3} - \tau_{231,3} = 0
\end{equation}
\begin{equation}
\label{3}
	\sigma_{31} - \tau_{311,1} - \tau_{313,3} - \tau_{331,3} = 0
\end{equation}
\begin{equation}
\label{4}
	\tau_{111} = 0, \quad \tau_{211} = 0, \quad \tau_{311} = 0
\end{equation}
\begin{equation}
\label{5}
	D_{1} - Q_{11,1} - Q_{13,3} - Q_{31,3} = 0
\end{equation}
\begin{equation}
\label{6}
	\hspace{-3.5cm}  Q_{11} = 0
\end{equation}
\vspace{-0.4cm} 
\begin{equation*}
	\hspace{-3.5cm} \underline{x_3 = \pm h}:
\end{equation*}
\begin{equation}
\label{7}
	\sigma_{13} - \tau_{113,1} - \tau_{131,1} - \tau_{133,3} = 0
\end{equation}
\begin{equation}
\label{8}
	\sigma_{23} - \tau_{213,1} - \tau_{231,1} - \tau_{233,3} = 0
\end{equation}
\begin{equation}
\label{9}
	\sigma_{33} - \tau_{313,1} - \tau_{331,1} - \tau_{333,3} = 0
\end{equation}
\begin{equation}
\label{10}
	\tau_{133} = 0, \quad \tau_{233} = 0, \quad \tau_{333} = 0
\end{equation}
\begin{equation}
\label{11}
	D_{3} - Q_{13,1} - Q_{31,1} - Q_{33,3} = 0
\end{equation}
\begin{equation}
\label{12}
	\hspace{-3.5cm}  Q_{33} = 0
\end{equation}

Following Saint-Venant's approach of classical elasticity, that was also used in electroelasticity (see, e.g. \cite{Davi1996}) and in strain gradient elasticity \cite{Lurie2018} we will use a relaxed formulation of boundary conditions on the end faces of the layer. Thus, we assume that stress resultants, rather than pointwise tractions, are prescribed there and we use the following conditions instead of (\ref{1}):
\begin{equation}
\label{N}
	\int_{-h}^{h} (\sigma_{11} - \tau_{111,1} - \tau_{113,3} - \tau_{131,3})
	dx_3 = 0
\end{equation}
\begin{equation}
\label{M}
	\int_{-h}^{h} x_3 (\sigma_{11} - \tau_{111,1} - \tau_{113,3} - \tau_{131,3}) 
	dx_3 = M
\end{equation}

Edge boundary conditions (\ref{Edge}) reduce to the following relations at the layer corners:
\begin{equation}
\label{Edgei}
\begin{aligned}
	&\underline{x_1 =0, L, \, x_3 = \pm h}:\\
	&\tau_{113} = -\tau_{131}, 
	\quad \tau_{213} = -\tau_{231}, 
	\quad \tau_{313} = -\tau_{331}, 
	\quad Q_{13} = -Q_{31}
\end{aligned}
\end{equation}

Next, let suggest the solution for the displacements and electric potential in the following form:
\begin{equation}
\label{Hyp}
	u_1 = K x_1 x_3, \qquad 
	u_3 = -\frac{1}{2}K x_1^2 + w(x_3), \qquad
	\phi = \varphi (x_3)
\end{equation}
where $K$, $w(x_3)$ and $\varphi(x_3)$ are the unknown constant and functions that define the deformations and electric field in the layer and that should be determined from the governing equations of the problem.

From (\ref{Hyp}) it follows that non-zero strains, strain gradients, electric field component and its gradient in the layer are:
\begin{equation}
\label{eE}
\begin{aligned}
	&\varepsilon_{11} = K x_3, \quad \varepsilon_{33} = w',  \\
	&\varepsilon_{11,3} = K, \quad \varepsilon_{33,3} = w'', \\
	&E_3 = - \varphi', \quad E_{3,3} = - \varphi''
\end{aligned}
\end{equation}
where we introduce notation: $f' = \partial f/\partial x_3$.

We can mention here that the physical meaning of the constant $K$ is a through-thickness gradient of axial strains, which has a constant value in given solution like in classical elasticity. Thus, like in classical elasticity, hypotheses (\ref{Hyp}) impose that there are no shear deformations under pure bending in the layer and that its cross sections remain plane after deformations. This is a common assumption that used in the  analogous problems in generalized continuum theories and it is valid far from the end faces of the layer (see discussion in \cite{Lurie2018, Cowin1983, lurie2018mams, dell1997porous}). In the case of transverse isotropy this hypothesis also remain valid due absence of coupling between shear and normal strain for the chosen direction of poling. 

Substituting (\ref{eE}) into constitutive equations (\ref{s})-(\ref{Q}) and taking into account transverse isotropy of the layer material in considered simplified gradient theory, one can find the following non-zero stresses, double-stresses, electric displacement and electric quadruple:
\begin{equation}
\label{sij}
\begin{aligned}
	&\sigma_{11} = C_{11} K x_3 +C_{13} w' + e_{31} \varphi',  \\
	&\sigma_{22} = C_{12} K x_3 +C_{13} w' + e_{31} \varphi',  \\
	&\sigma_{33} = C_{13} K x_3 +C_{33} w' + e_{33} \varphi'
\end{aligned}
\end{equation}
\begin{equation}
\label{tauijk}
\begin{aligned}
	&\tau_{113} = \ell^2(C_{11} K + C_{13} w'' )\\
	&\tau_{223} = \ell^2(C_{12} K + C_{13} w'' )\\
	&\tau_{333} = \ell^2(C_{13} K + C_{33} w'')
\end{aligned}
\end{equation}
\begin{equation}
\label{Di}
\begin{aligned}
	&D_3 = e_{31} K x_3 + e_{33} w' - \kappa_{33}\varphi'
\end{aligned}
\end{equation}
\begin{equation}
\label{Qi}
\begin{aligned}
	&Q_{33} = - \ell^2 \kappa_{33}\varphi''
\end{aligned}
\end{equation}
where the Voigt notation is used for the material constants.

One can see that proposed solution (\ref{Hyp})-(\ref{Qi}) satisfy boundary conditions (\ref{2})--(\ref{8}), (\ref{10}.1), (\ref{10}.2), (\ref{Edgei}.2), (\ref{Edgei}.3). Equilibrium equations (\ref{Equil}) and generalized Gauss' law (\ref{Gauss}) in the transversal $x_2$ and longitudinal $x_1$ directions are also satisfied identically. Boundary conditions at the edges (\ref{Edgei}.1) can not be satisfied with the considered solution because the double stress $\tau_{113}$ does not equal to zero (see (\ref{tauijk})), while $\tau_{131}$ vanishes. However, this edge condition can be ignored in the \textit{Saint-Venant's} sense because its influence will be significant only near to the layer end faces. For example, these edge conditions should not affect the apparent bending stiffness of the layer. In strain gradient elasticity, it was shown by using numerical simulations in \cite{Lurie2018}. 

Thus, our task now is to solve the governing equations (\ref{Equil}), (\ref{Gauss}) in the layer thickness direction with boundary conditions on the top/bottom surfaces (\ref{9}), (\ref{10}.3), (\ref{11}), (\ref{12}) and satisfy the last relaxed boundary conditions at the beam end faces (\ref{M}), (\ref{N}). In terms of introducing unknown functions we have the following statement of the boundary value problem:
\begin{equation}
\label{BVP1}
\begin{aligned}
	&\underline{x_3\in[-h,h]}:\\
	&\ell^2 \,C_{33}w^{IV} - C_{33}w''-e_{33} \varphi'' - K C_{13} = 0,\\
	&\ell^2 \,\kappa_{33} \varphi^{IV} - \kappa_{33}\varphi''+e_{33} w'' + K e_{31} = 0
\end{aligned}
\end{equation}
\begin{equation}
\label{BVP2}
\begin{aligned}
	&\underline{x_3=\pm h}:\\
	&\ell^2 \,C_{33}w''' - C_{33}w'-e_{33} \varphi' - K C_{13} x_3= 0,\\
	&\ell^2 \,\kappa_{33}\varphi''' - \kappa_{33}\varphi'+e_{33} w' + K e_{31}x_3 = 0,\\
	&C_{33} w'' + K C_{13} = 0\\
	&\varphi''= 0
\end{aligned}
\end{equation}

General solution of the system (\ref{BVP1}) can be represented in the following form accounting for the symmetry of boundary conditions (\ref{BVP2}):
\begin{equation}
\label{solwf}
\begin{aligned}
	\varphi(x_3) = \,&
	K \frac{(C_{33} e_{31}-C_{13} e_{33})}
	{2 (C_{33} \kappa_{33}+e_{33}^2)} x_3^2 + A_{01} +
	\sum_{i=1}^2 A_i \cosh{\frac{\lambda_i x_3}{\ell}},\\
	w(x_3) = \,&- K \frac{(e_{33} e_{31}+C_{13} \kappa_{33})}
	{2 (C_{33} \kappa_{33}+e_{33}^2)} x_3^2 + A_{02} +
	\frac{\kappa_{33}}{e_{33}}\sum_{i=1}^2 A_i (1-\lambda_i^2)
	\cosh{\frac{\lambda_i x_3}{\ell}},
\end{aligned}
\end{equation}
where $k_{33}= e_{33}/\sqrt{C_{33}\kappa_{33}}$ is the electromechanical coupling factor of the layer material and $\lambda_{1,2} = \sqrt{1\pm i \,k_{33}}$ are the eigenvalues of ODE system ({\ref{BVP1}).

Constants $A_{01}$, $A_{02}$, $A_1$, $A_2$ in (\ref{solwf}) should be found from the boundary conditions on the top or bottom surface of the beam (\ref{BVP2}) taking into account fixed condition for displacement and zero value of electric potential at the origin of coordinate system, i.e. $w(0) = 0$ and $\varphi(0)=0$. Solutions for this constants are the following:
\begin{equation}
\label{Ai}
\begin{aligned}
	&A_{01} = - S \left ( \frac{(k_{33}-i)^2}{\cosh \overline{h}_1} +
	\frac{(k_{33}+i)^2}{\cosh \overline{h}_2} \right)\\
	&A_{02} = - S \sqrt{\frac{\kappa_{33}}{C_{33}}} 
	\left ( \frac{i (k_{33}-i)^2}{\cosh \overline{h}_1} -
	\frac{i (k_{33}+i)^2}{\cosh \overline{h}_2} \right)\\
	&A_{1} = S \frac{(k_{33}+i)^2}{\cosh \overline{h}_2}, \quad
	A_{2} = S \frac{(k_{33}-i)^2}{\cosh \overline{h}_1}\\
\end{aligned}
\end{equation}
where we use the notations:
\begin{equation*}
\label{eE}
\begin{aligned}
	&S = K \ell^2 \frac{C_{33} e_{31} - C_{13} e_{33}}
	{(1+k_{33}^2)^2 C_{33} \kappa_{33}}, \\
	&\overline{h}_1 =\,\frac{\lambda_2 h}{\ell}, \quad
	\overline{h}_2 = \frac{\lambda_1 h}{\ell}
\end{aligned}
\end{equation*}

It can be checked, that despite the fact that the eigenvalues $\lambda_i$ are imagine in solution (\ref{solwf}), the values of functions $w(x_3)$ and $\varphi(x_3)$ will be always real for any physically admissible values of materials constants and geometrical parameters of the problem. These functions are found now up to the constant $K$ that should be found by using relaxed boundary conditions at the layer end faces. To do this, we should substitute (\ref{solwf}), (\ref{Ai}) into (\ref{u}), (\ref{E}) and then into (\ref{sij}), (\ref{tauijk}) to find the stresses and double stresses. After that one can evaluate the integrals in boundary conditions (\ref{N}), (\ref{M}). It can be shown that the first condition (\ref{N}) for the force resultants will be satisfied identically. The use of a bending moments condition (\ref{M}) provides us the following result:
\begin{equation}
\label{K}
	K = \frac{M}{D J}\\
\end{equation}
where $D= E^* I$ is classical apparent bending stiffness of the piezoelectric layer under pure bending, which is realized in the absence of gradient effects, i.e. if $\ell=0$; $I= 2h^3/3$ is the moment of inertia per unit width of the layer and $E^*$ is classical apparent Young's modulus of piezoelectric layer that is defined by the relation:
\begin{equation*}
\label{Eeff}
	E^* = \frac{C_{33} e_{31}^2 - 2 C_{13} e_{31} e_{33} + C_{11} 		e_{33}^2 - C_{13}^2 \kappa_{33} + C_{11}C_{33}\kappa_{33}}
	{e_{33}^2 + C_{33} \kappa_{33}}
\end{equation*}
Non-classical effects in (\ref{K}) are defined by the non-dimensional function $J$ that can be presented in the following form:
\begin{equation}
\label{J}
\begin{aligned}
	J = \,&1+ 
	\frac{\ell^2}{h^2}
	\frac{3 \kappa_{33} (C_{33} e_{31} - C_{13} e_{33})^2}
	{2 E^* (e_{33}^2 + C_{33} \kappa_{33})^2}
	\sum_{i=1}^2 \left
	(1+k_{33}^2 - 2\lambda_i^2\right)
	\frac{\tanh{\overline{h}_i}}{\overline{h}_i}
\end{aligned}
\end{equation}

Thus, we derived a closed-form semi-inverse analytical solution (\ref{solwf})-(\ref{J}) for a pure bending problem of the piezoelectric layer taking into account electric field and strain gradient effects. This solution can be reduced to the classical one (see e.g. \cite{Lim2001}) by assuming zero value of the length scale parameter $\ell = 0$. In absence of electromechanical coupling ($e_{33}$ = $e_{31}$ = 0) this solution becomes to the corresponding solution of the strain gradient elasticity theory \cite{Lurie2018}. In absence both of the gradient and electromechanical effects the solution coincides with famous classical elasticity result, namely, the constant, which define the linear variation of axial strain across the layer thickness, is defined by relation $K = M/(E^* I)$, where $E^* = E_{1} = C_{11} - \frac{C_{13}^2}{C_{33}}$ is an elastic modulus of the layer material in $x_1$ direction.

\section{Analytical solution for the thin piezoelectric layer}
\label{sec:5}

Now, we can use obtained semi-inverse solution to derive an asymptotic solution for a layer of very small thickness. In considered gradient model we should assume then that the ratio $h/\ell$ is very small and tend to zero. It can be shown, that in this limiting case the solution (\ref{solwf})-(\ref{Ai}),  reduces to the classical one and does not depend on the length scale parameter. Namely, the deflections of the layer neutral axis (at $x_3=0$) and electric potential function can be found in the following form:
\begin{equation}
\label{thin}
	u_3 = -\frac{M}{2E_1 I}x_1^2, 
	\quad\quad
	\phi = \frac{M}{2E_1 I}
	\frac{C_{33} e_{31}-C_{13} e_{33}}{e_{33}^2 + C_{33} \kappa_{33}} x_3^2
\end{equation}
where we take into account that $w|_{x_3=0}=0$ and that $\cosh\tfrac{\lambda_i z}{\ell} \approx 1$ if $z\in[-h,h]$ and $h/\ell \rightarrow 0$.

Obtained simple result can be used for verification of the Kirchoff plate and Euler-Bernoulli beam models (taking into account the relations between elastic constants in the plane strain and plane stress statements of the theory) in second gradient electroelasticity. One can see, that the electromechanical response of thin structures under pure bending should be classical according to (\ref{thin}). 

Note, that similar solutions have been used recently to verify the beam theories in the strain gradient and distortion gradient elasticity theories (Mindlin Form I and II) in \cite{Lurie2018} and in the framework of micro-dilatation elasticity in \cite{Birsan2011}. It was shown that the apparent bending stiffness of the thin beams in these theories should have classical values. The same effects we obtain in the second gradient electroelasticity -- apparent bending stiffness and voltage of gradient piezoelectric plates and beams are classical. 

\section{Conclusions}
\label{sec:5}
Semi-inverse analytical solution is developed for a pure bending problem in the framework of plane strain statement of the second gradient electroelasticity theory. Obtained solution is valid far from the layer end faces where the shear stress and corresponding distortions arise. Numerical finite-element validation of this fact is given in the forthcoming article by the authors.

Presented solution can be used as the test solution for the plate and beam models in considered gradient theory. It is shown that the electromechanical response of thin layer under pure bending will correspond to the classical solution without arising non-classical size effects. Additionally, it is proved that the size-dependent behavior of thin piezoelectric structures that were found in experiments cannot be related to the strain and electric filed gradient effects. Another model should be involved to describe such phenomena, for example, taking into account surface or micropolar effects. 

\bibliographystyle{unsrtnat}
\bibliography{refs}

\begin{thebibliography}{56}
\providecommand{\natexlab}[1]{#1}
\providecommand{\url}[1]{\texttt{#1}}
\expandafter\ifx\csname urlstyle\endcsname\relax
  \providecommand{\doi}[1]{doi: #1}\else
  \providecommand{\doi}{doi: \begingroup \urlstyle{rm}\Url}\fi

\bibitem[Yang(2006)]{Yang2006}
Jiashi Yang.
\newblock {A Review of a Few Topics in Piezoelectricity}.
\newblock \emph{Applied Mechanics Reviews}, 59\penalty0 (6):\penalty0 335,
  2006.
\newblock ISSN 00036900.
\newblock \doi{10.1115/1.2345378}.

\bibitem[Majdoub et~al.(2008)Majdoub, Sharma, and
  {\c{C}}a{\u{g}}in]{majdoub2008}
MS~Majdoub, P~Sharma, and T~{\c{C}}a{\u{g}}in.
\newblock Dramatic enhancement in energy harvesting for a narrow range of
  dimensions in piezoelectric nanostructures.
\newblock \emph{Physical Review B}, 78\penalty0 (12):\penalty0 121407, 2008.

\bibitem[Yan and Jiang(2017)]{Yan2017}
Zhi Yan and Liying Jiang.
\newblock {Modified Continuum Mechanics Modeling on Size-Dependent Properties
  of Piezoelectric Nanomaterials : A Review}.
\newblock pages 1--18, 2017.
\newblock \doi{10.3390/nano7020027}.

\bibitem[Oates(2017)]{Oates2017}
William~S. Oates.
\newblock {Flexoelectricity, strain gradients, and singularities in
  ferroelectric nanostructures}.
\newblock \emph{Journal of Intelligent Material Systems and Structures},
  28\penalty0 (20):\penalty0 3091--3105, 2017.
\newblock ISSN 15308138.
\newblock \doi{10.1177/1045389X17704985}.

\bibitem[Sharma et~al.(2007)Sharma, Maranganti, and Sharma]{Sharma2007}
N~D Sharma, R~Maranganti, and P~{\~{A}} Sharma.
\newblock {On the possibility of piezoelectric nanocomposites without using
  piezoelectric materials}.
\newblock 55:\penalty0 2328--2350, 2007.
\newblock \doi{10.1016/j.jmps.2007.03.016}.

\bibitem[{Tagantsev. P V} and Yudin(2013)]{Tagantsev2013}
{Tagantsev. P V} and A~K Yudin.
\newblock {Fundamentals of flexoelectricity in solids}.
\newblock \emph{Nanotechnology}, 24\penalty0 (43), 2013.
\newblock ISSN 09574484.
\newblock \doi{10.1088/0957-4484/24/43/432001}.

\bibitem[Wang and Wang(2018{\natexlab{a}})]{Wang2018}
K.F. Wang and B.L. Wang.
\newblock {Electrostatic potential in a bent piezoelectric nanowire with
  consideration of size-dependent piezoelectricity and semiconducting
  characterization}.
\newblock pages 0--18, 2018{\natexlab{a}}.
\newblock ISSN 2190-4995.
\newblock \doi{https://doi.org/10.1088/1478-3975/aa9768}.

\bibitem[Mindlin(1968)]{mindlin1968}
Raymond~David Mindlin.
\newblock Polarization gradient in elastic dielectrics.
\newblock \emph{International Journal of Solids and Structures}, 4\penalty0
  (6):\penalty0 637--642, 1968.

\bibitem[Sahin and Dost(1988)]{sahin1988}
E~Sahin and S~Dost.
\newblock A strain-gradients theory of elastic dielectrics with spatial
  dispersion.
\newblock \emph{International journal of engineering science}, 26\penalty0
  (12):\penalty0 1231--1245, 1988.

\bibitem[Kafadar(1971)]{kafadar1971}
Chas~B Kafadar.
\newblock The theory of multipoles in classical electromagnetism.
\newblock \emph{International Journal of Engineering Science}, 9\penalty0
  (9):\penalty0 831--853, 1971.

\bibitem[Arvanitakis(2018)]{Arvanitakis2018}
Antonios Arvanitakis.
\newblock {Gradient effects in a new class of electro-elastic bodies}.
\newblock \emph{Zeitschrift f{\"{u}}r angewandte Mathematik und Physik},
  69\penalty0 (3):\penalty0 62, 2018.
\newblock ISSN 0044-2275.
\newblock \doi{10.1007/s00033-018-0959-0}.
\newblock URL \url{http://link.springer.com/10.1007/s00033-018-0959-0}.

\bibitem[Liang and Shen(2013)]{Liang2013x}
Xu~Liang and Shengping Shen.
\newblock {Size-Dependent Piezoelectricity and Elasticity Due To the Electric
  Field-Strain Gradient Coupling and Strain Gradient Elasticity}.
\newblock \emph{International Journal of Applied Mechanics}, 05\penalty0
  (02):\penalty0 1350015, 2013.
\newblock ISSN 1758-8251.
\newblock \doi{10.1142/S1758825113500154}.
\newblock URL
  \url{http://www.worldscientific.com/doi/abs/10.1142/S1758825113500154}.

\bibitem[Enakoutsa et~al.(2015)Enakoutsa, Vescovo, and Scerrato]{Enakoutsa2015}
K~Enakoutsa, D~Del Vescovo, and D~Scerrato.
\newblock {Combined polarization field gradient and strain field gradient
  effects in elastic flexoelectric materials}.
\newblock 2015.
\newblock \doi{10.1177/1081286515616048}.

\bibitem[Ie?an(2018)]{Iesan2018}
D.~Ie?an.
\newblock {A theory of thermopiezoelectricity with strain gradient and electric
  field gradient effects}.
\newblock \emph{European Journal of Mechanics, A/Solids}, 67:\penalty0
  280--290, 2018.
\newblock ISSN 09977538.
\newblock \doi{10.1016/j.euromechsol.2017.09.007}.

\bibitem[Liu et~al.(2016)Liu, Ke, Yang, Kitipornchai, and Wang]{Liu2016}
Chen Liu, Liao-liang Ke, Jie Yang, Sritawat Kitipornchai, and Yue-sheng Wang.
\newblock {Buckling and post-buckling analyses of size-dependent piezoelectric
  nanoplates}.
\newblock \emph{Theoretical {\&} Applied Mechanics Letters}, 6\penalty0
  (6):\penalty0 253--267, 2016.
\newblock ISSN 2095-0349.
\newblock \doi{10.1016/j.taml.2016.10.003}.
\newblock URL \url{http://dx.doi.org/10.1016/j.taml.2016.10.003}.

\bibitem[Hadjesfandiari(2013)]{Hadjesfandiari2013}
Ali~R. Hadjesfandiari.
\newblock {Size-dependent piezoelectricity}.
\newblock \emph{International Journal of Solids and Structures}, 50\penalty0
  (18):\penalty0 2781--2791, 2013.
\newblock ISSN 00207683.
\newblock \doi{10.1016/j.ijsolstr.2013.04.020}.
\newblock URL \url{http://dx.doi.org/10.1016/j.ijsolstr.2013.04.020}.

\bibitem[Malikan(2017)]{Malikan2017}
Mohammad Malikan.
\newblock {Electro-mechanical shear buckling of piezoelectric nanoplate using
  modified couple stress theory based on simplified first order shear
  deformation theory}.
\newblock \emph{Applied Mathematical Modelling}, 48:\penalty0 196--207, 2017.
\newblock ISSN 0307904X.
\newblock \doi{10.1016/j.apm.2017.03.065}.
\newblock URL \url{http://dx.doi.org/10.1016/j.apm.2017.03.065}.

\bibitem[Liang et~al.(2014)Liang, Hu, and Shen]{liang2014}
Xu~Liang, Shuling Hu, and Shengping Shen.
\newblock Effects of surface and flexoelectricity on a piezoelectric nanobeam.
\newblock \emph{Smart Materials and Structures}, 23\penalty0 (3):\penalty0
  035020, 2014.

\bibitem[Shen and Hu(2010)]{shen2010}
Shengping Shen and Shuling Hu.
\newblock A theory of flexoelectricity with surface effect for elastic
  dielectrics.
\newblock \emph{Journal of the Mechanics and Physics of Solids}, 58\penalty0
  (5):\penalty0 665--677, 2010.

\bibitem[Wang and Wang(2012)]{Wang2012}
K.~F. Wang and B.~L. Wang.
\newblock {The electromechanical coupling behavior of piezoelectric nanowires:
  Surface and small-scale effects}.
\newblock \emph{Epl}, 97\penalty0 (6), 2012.
\newblock ISSN 02955075.
\newblock \doi{10.1209/0295-5075/97/66005}.

\bibitem[Liang et~al.(2013)Liang, Hu, and Shen]{liang2013}
Xu~Liang, Shuling Hu, and Shengping Shen.
\newblock Bernoulli--euler dielectric beam model based on strain-gradient
  effect.
\newblock \emph{Journal of Applied Mechanics}, 80\penalty0 (4):\penalty0
  044502, 2013.

\bibitem[Wang and Wang(2018{\natexlab{b}})]{wang2018b}
KF~Wang and BL~Wang.
\newblock Electrostatic potential in a bent piezoelectric nanowire with
  consideration of size-dependent piezoelectricity and semiconducting
  characterization.
\newblock \emph{Nanotechnology}, 29\penalty0 (25):\penalty0 255405,
  2018{\natexlab{b}}.

\bibitem[Yue et~al.(2015)Yue, Xu, and Aifantis]{yue2015b}
Yanmei Yue, Kaiyu Xu, and Elias~C Aifantis.
\newblock Strain gradient and electric field gradient effects in piezoelectric
  cantilever beams.
\newblock \emph{Journal of the Mechanical Behavior of Materials}, 24\penalty0
  (3-4):\penalty0 121--127, 2015.

\bibitem[Yang et~al.(2015)Yang, Liang, and Shen]{yang2015b}
Wenjun Yang, Xu~Liang, and Shengping Shen.
\newblock Electromechanical responses of piezoelectric nanoplates with
  flexoelectricity.
\newblock \emph{Acta Mechanica}, 226\penalty0 (9):\penalty0 3097--3110, 2015.

\bibitem[Baroudi et~al.(2018)Baroudi, Najar, and Jemai]{Baroudi2018}
S~Baroudi, F~Najar, and A~Jemai.
\newblock {International Journal of Solids and Structures Static and dynamic
  analytical coupled field analysis of piezoelectric flexoelectric nanobeams :
  A strain gradient theory approach}.
\newblock \emph{International Journal of Solids and Structures}, 135:\penalty0
  110--124, 2018.
\newblock ISSN 0020-7683.
\newblock \doi{10.1016/j.ijsolstr.2017.11.014}.
\newblock URL \url{https://doi.org/10.1016/j.ijsolstr.2017.11.014}.

\bibitem[Lurie and Belyaev(2005)]{lurie2005}
AI~Lurie and A~Belyaev.
\newblock Theory of elasticity. foundations of engineering mechanics.
\newblock \emph{Springer, Berlin}, 10\penalty0 (1007):\penalty0 978--3, 2005.

\bibitem[Iesan(1989)]{Iesan1989}
D~Iesan.
\newblock {On Saint-Venant's problem for elastic dielectrics}.
\newblock \emph{Journal of Elasticity 21:}, 32:\penalty0 101--115, 1989.
\newblock \doi{10.1115/1.1365152}.

\bibitem[Davi(1996)]{Davi1996}
F~Davi.
\newblock {Saint-Venant's problem for linear piezoelectric bodies}.
\newblock \emph{Journal of Elasticity}, 43\penalty0 (3):\penalty0 227--245,
  1996.
\newblock ISSN 0374-3535.
\newblock \doi{10.1007/BF00042502}.

\bibitem[Dell'Isola and Rosa(1996)]{DellIsola1996}
Francesco Dell'Isola and Luigi Rosa.
\newblock {Saint Venant problem in linear piezoelectricity}.
\newblock \emph{In Smart Structures and Materials 1996: Mathematics and Control
  in Smart Structures, SPIE}, 2715\penalty0 (2):\penalty0 399--410, 1996.

\bibitem[Bisegna(1998)]{Bisegna1998}
P~Bisegna.
\newblock {The Saint-Venant Problem for Monoclinic Piezoelectric Cylinders}.
\newblock \emph{Z. Angew. Math. Mech.}, 78\penalty0 (3):\penalty0 147--165,
  1998.

\bibitem[Rovenski et~al.(2007)Rovenski, Harash, and Abramovich]{Rovenski2007}
Vladimir Rovenski, Eugene Harash, and Haim Abramovich.
\newblock {Staint-Venant 's Problem for Homogeneous Piezoelectric Beams}.
\newblock \emph{Journal of Applied Mechanics}, 74\penalty0 (July):\penalty0
  1095--1103, 2007.
\newblock ISSN 00218936.
\newblock \doi{10.1115/1.2722315}.

\bibitem[Krommer(2001)]{Krommer2001}
M.~Krommer.
\newblock {On the correction of the Bernoulli-Euler beam theory for smart
  piezoelectric beams}.
\newblock \emph{Smart Materials and Structures}, 10\penalty0 (4):\penalty0
  668--680, 2001.
\newblock ISSN 09641726.
\newblock \doi{10.1088/0964-1726/10/4/310}.

\bibitem[Heyliger and Brooks(1996)]{Heyliger1996}
P.~R. Heyliger and S.~Brooks.
\newblock {Exact Solutions for Laminated Piezoelectric Plates in Cylindrical
  Bending}.
\newblock \emph{Journal of Applied Mechanics}, 63\penalty0 (4):\penalty0
  903--910, 1996.
\newblock ISSN 00218936.
\newblock \doi{10.1115/1.2787245}.
\newblock URL \url{http://dx.doi.org/10.1115/1.2787245}.

\bibitem[Heyliger(1997)]{Heyliger1997}
P.~Heyliger.
\newblock {Exact Solutions for Simply Supported Laminated Piezoelectric
  Plates}.
\newblock \emph{Journal of Applied Mechanics}, 64\penalty0 (2):\penalty0 299,
  1997.
\newblock ISSN 00218936.
\newblock \doi{10.1115/1.2787307}.

\bibitem[Cowin and Nunziato(1983)]{Cowin1983}
Stephen~C. Cowin and Jace~W. Nunziato.
\newblock {Linear elastic materials with voids}.
\newblock \emph{Journal of Elasticity}, 13\penalty0 (2):\penalty0 125--147,
  1983.
\newblock ISSN 03743535.
\newblock \doi{10.1007/BF00041230}.

\bibitem[B{\^{i}}rsan and Altenbach(2011)]{Birsan2011}
Mircea B{\^{i}}rsan and Holm Altenbach.
\newblock {On the theory of porous elastic rods}.
\newblock \emph{International Journal of Solids and Structures}, 48\penalty0
  (6):\penalty0 910--924, 2011.
\newblock ISSN 00207683.
\newblock \doi{10.1016/j.ijsolstr.2010.11.022}.

\bibitem[Ie{\c{s}}an(1971)]{iesan1971}
Dorin Ie{\c{s}}an.
\newblock On saint-venant's problem in micropolar elasticity.
\newblock \emph{International Journal of Engineering Science}, 9\penalty0
  (10):\penalty0 879--888, 1971.

\bibitem[Reddy and Venkatasubramanian(1978)]{reddy1978}
GV~Krishna Reddy and NK~Venkatasubramanian.
\newblock On the flexural rigidity of a micropolar elastic circular cylinder.
\newblock \emph{Journal of Applied Mechanics}, 45\penalty0 (2):\penalty0
  429--431, 1978.

\bibitem[Iesan and Nappa(1994)]{iesan1994}
D~Iesan and L~Nappa.
\newblock Saint-venants problem for microstretch elastic solids.
\newblock \emph{International journal of engineering science}, 32\penalty0
  (2):\penalty0 229--236, 1994.

\bibitem[Lurie and Solyaev(2018)]{Lurie2018}
S.~Lurie and Y.~Solyaev.
\newblock {Revisiting bending theories of elastic gradient beams}.
\newblock \emph{International Journal of Engineering Science}, 126:\penalty0
  1--21, 2018.
\newblock ISSN 00207225.
\newblock \doi{10.1016/j.ijengsci.2018.01.002}.
\newblock URL \url{https://doi.org/10.1016/j.ijengsci.2018.01.002}.

\bibitem[Lurie et~al.(2016)Lurie, Volkov-Bogorodsky, Belov, and
  Lykosova]{lurie2016nanosized}
Sergey~A Lurie, Dmitriy~B Volkov-Bogorodsky, PA~Belov, and ED~Lykosova.
\newblock Do nanosized rods have abnormal mechanical properties? on some
  fallacious ideas and direct errors related to the use of the gradient
  theories for simulation of scale-dependent rods.
\newblock \emph{Nanoscience and Technology: An International Journal},
  7\penalty0 (4):\penalty0 261--295, 2016.

\bibitem[Mindlin(1964)]{Mindlin1964}
R.~D. Mindlin.
\newblock {Micro-structure in linear elasticity}.
\newblock \emph{Archive for Rational Mechanics and Analysis}, 16\penalty0
  (1):\penalty0 51--78, 1964.
\newblock ISSN 00039527.
\newblock \doi{10.1007/BF00248490}.

\bibitem[Mindlin and Eshel(1968)]{mindlin1968b}
Raymond~David Mindlin and NN~Eshel.
\newblock On first strain-gradient theories in linear elasticity.
\newblock \emph{International Journal of Solids and Structures}, 4\penalty0
  (1):\penalty0 109--124, 1968.

\bibitem[Kalpakides and Agiasofitou(2002)]{Kalpakides2002}
V.~K. Kalpakides and E.~K. Agiasofitou.
\newblock {On material equations in second gradient electroelasticity}.
\newblock \emph{Journal of Elasticity}, 67\penalty0 (3):\penalty0 205--227,
  2002.
\newblock ISSN 03743535.
\newblock \doi{10.1023/A:1024926609083}.

\bibitem[Hu and Shen(2009)]{hu2009}
Shuling Hu and Shengping Shen.
\newblock Electric field gradient theory with surface effect for
  nano-dielectrics.
\newblock \emph{Computers, Materials \& Continua (CMC)}, 13\penalty0
  (1):\penalty0 63, 2009.

\bibitem[Sladek et~al.(2018)Sladek, Sladek, W{\"{u}}nsche, and
  Zhang]{Sladek2018}
Jan Sladek, Vladimir Sladek, Michael W{\"{u}}nsche, and Chuanzeng Zhang.
\newblock {Effects of electric field and strain gradients on cracks in
  piezoelectric solids}.
\newblock \emph{European Journal of Mechanics - A/Solids}, 71\penalty0 (June
  2017):\penalty0 187--198, 2018.
\newblock ISSN 09977538.
\newblock \doi{10.1016/j.euromechsol.2018.03.018}.
\newblock URL
  \url{http://linkinghub.elsevier.com/retrieve/pii/S0997753817304473}.

\bibitem[Yue et~al.(2014)Yue, Xu, and Aifantis]{Yue2014}
Y.~M. Yue, K.~Y. Xu, and E.~C. Aifantis.
\newblock {Microscale size effects on the electromechanical coupling in
  piezoelectric material for anti-plane problem}.
\newblock \emph{Smart Materials and Structures}, 23\penalty0 (12), 2014.
\newblock ISSN 1361665X.
\newblock \doi{10.1088/0964-1726/23/12/125043}.

\bibitem[Solyaev and Lurie(2018)]{Solyaev2018}
Y.~Solyaev and S.~Lurie.
\newblock {Numerical predictions for the effective size-dependent properties of
  piezoelectric composites with spherical inclusions}.
\newblock \emph{Composite Structures}, 202\penalty0 (January):\penalty0
  1099--1108, 2018.
\newblock ISSN 02638223.
\newblock \doi{10.1016/j.compstruct.2018.05.050}.
\newblock URL \url{https://doi.org/10.1016/j.compstruct.2018.05.050}.

\bibitem[Askes and Aifantis(2011)]{askes2011}
Harm Askes and Elias~C Aifantis.
\newblock Gradient elasticity in statics and dynamics: an overview of
  formulations, length scale identification procedures, finite element
  implementations and new results.
\newblock \emph{International Journal of Solids and Structures}, 48\penalty0
  (13):\penalty0 1962--1990, 2011.

\bibitem[Park and Gao(2006)]{park2006}
SK~Park and XL~Gao.
\newblock Bernoulli--euler beam model based on a modified couple stress theory.
\newblock \emph{Journal of Micromechanics and Microengineering}, 16\penalty0
  (11):\penalty0 2355, 2006.

\bibitem[Eremeyev et~al.(2017)Eremeyev, Dell'Isola, Boutin, and
  Steigmann]{eremeyev2017}
Victor~A Eremeyev, Francesco Dell'Isola, Claude Boutin, and David Steigmann.
\newblock Linear pantographic sheets: existence and uniqueness of weak
  solutions.
\newblock \emph{Journal of Elasticity}, pages 1--22, 2017.

\bibitem[Parton and Kudryavtsev(1988)]{parton1988}
VZ~Parton and BA~Kudryavtsev.
\newblock Electromagnetoelasticity.
\newblock \emph{Gordon and Breach Science Publishers, New York}, 2:\penalty0
  90059--0, 1988.

\bibitem[Yang et~al.(2004)Yang, Hu, and Yang]{Yang2004}
X~M Yang, Y~T Hu, and J~S Yang.
\newblock {Electric field gradient effects in anti-plane problems of polarized
  ceramics}.
\newblock 41:\penalty0 6801--6811, 2004.
\newblock \doi{10.1016/j.ijsolstr.2004.05.018}.

\bibitem[Lurie et~al.(2018)Lurie, Solyaev, Volkov, and
  Volkov-Bogorodskiy]{lurie2018mams}
Sergey Lurie, Yury Solyaev, Alexander Volkov, and Dmitriy Volkov-Bogorodskiy.
\newblock Bending problems in the theory of elastic materials with voids and
  surface effects.
\newblock \emph{Mathematics and Mechanics of Solids}, 23\penalty0 (5):\penalty0
  787--804, 2018.

\bibitem[Dell'Isola and Batra(1997)]{dell1997porous}
Francesco Dell'Isola and Romesh~C Batra.
\newblock Saint-venant's problem for porous linear elastic materials.
\newblock \emph{Journal of elasticity}, 47\penalty0 (1):\penalty0 73--81, 1997.

\bibitem[Lim and He(2001)]{Lim2001}
C.~W. Lim and L.~H. He.
\newblock {Exact solution of a compositionally graded piezoelectric layer under
  uniform stretch, bending and twisting}.
\newblock \emph{International Journal of Mechanical Sciences}, 43\penalty0
  (11):\penalty0 2479--2492, 2001.
\newblock ISSN 00207403.
\newblock \doi{10.1016/S0020-7403(01)00059-5}.

\end{thebibliography}

\end{document}